\documentclass
[preprint,tightenlines,eqsecnum,aps,epsfig]{revtex4}

\usepackage{graphicx}
\usepackage{epstopdf}
\usepackage{epsfig}

\def\beq{\begin{equation}}   \def\eeq{
\end{equation}}

\begin{document}
\title{On the shape  of a rapid hadron in QCD. }
\author{ B. Blok\email{Email :blok@physics.technion.ac.il} }
\affiliation{Department of Physics, Technion---Israel Institute of Technology, 32000 Haifa, Israel}
\author{ L. Frankfurt\email{E-mail: frankfur@tauphy.tau.ac.il} }
\affiliation{School of Physics and Astronomy, Raymond and Beverly
Sackler Faculty of Exact Sciences, Tel Aviv University, 69978 Tel
Aviv, Israel}
\author{ M.Strikman\email{\E-mail: strikman@phys.psu.edu}}
\affiliation{Physics Department, Penn State University, University Park, PA, USA}
\thispagestyle{empty}
\begin{abstract}

We visualize the fundamental  property of pQCD: the smaller size
of the colorless quark-gluon configurations leads to a  more rapid
increase of its interaction with energy. Within the frame of the
dipole model we use the $k_t$ factorization theorem to generalize
the DGLAP approximation and/or leading $\ln(x_0/x)$ approximation
and evaluate  the interaction of the quark dipole with a target.
In the limit of  fixed $Q^2$ and $x\to 0$ we find the increase
with energy of transverse momenta of quark(antiquark) within the
q$\bar q$ pair produced by the strongly virtual photon. The
average $p^2_t$ is evaluated analytically within the double
logarithmic approximation.  We demonstrate  that the invariant
mass$^2$ of the q$\bar q$ pair increases with the energy as
$M^2_0(x_0/x)^{\lambda}$, where $\lambda\sim 0.4\alpha_sN_c/\pi$
for transverse photons, and  as $\sim M^2_0
\exp{0.17[(4\alpha_sN_c/\pi)\log(x_0/x)]^{1/2}}$ for longitudinal
photons, where $M^2_0 \approx 0.7Q^2$ at the energies of the order
$s_0\sim 10^4$ GeV$^2$ ($x_0\sim 10^{-2}$. The magnitude of the
effect depends strongly on the small $x$ behavior of the gluon
distribution. Similar pattern of the energy dependence of $M^2$ is
found in the LO DGLAP approximation generalized to account for the
$k_t$ factorization.  We discuss the impact of the found
phenomenon on the dependence of the coherence length on the
initial energy  and demonstrate that the shape of the final hadron
state in DIS   has the biconcave form instead of the pancake. Some
implications of the found phenomena for the hard processes in pp
collisions are discussed.
\end{abstract}

\maketitle \setcounter{page}{1}
\section{Introduction.}
\par
A dipole model developed in ref. \cite{BFGMS}, cf.also \cite{AFS,
FMS,BBFS,Muellereikonal,CFS}  is the generalization of the parton
model  to the target rest frame description. It accounts for the
effects of the $Q^2$ and $\ln(x_0/x)$ evolutions. It also provides
the solution of the equations of QCD in the kinematics of fixed
and not too small $x=Q^2/\nu$ but $Q^2\to \infty$. The
characteristic feature of this solution  is the approximate
Bjorken  scaling for the structure functions of DIS, i.e. the two
dimensional conformal invariance for the moments of the structure
functions. In this approximation as well as within the leading
$\ln(x_0/x)$ approximation, the transverse momenta of quarks
within the dipole produced by the local electroweak current are
restricted by the virtuality of the external field:
\begin{equation}
\lambda_{QCD}^2 \le p_t^2 \le (Q^2)/4.
\end{equation}
  The aim of the present  paper is to demonstrate  that
the transverse  momenta of (anti)quark of the   $q\bar q$ pair
produced by a local current also increase  with energy and become
larger than $Q^2$ at sufficiently large energies. Technically this
effect follows from the more  rapid increase with the energy of
the pQCD interaction
 for smaller dipole and $k_t$ factorization.
 Let us note that this phenomena is different from the
 well known Lipatov diffusion. The latter means that within the
  leading log $\alpha_s\ln(x_0/x)$ approximation  the
parton transverse momenta are increasing with energy in the center
of rapidity as $\ln^2(p_{t}^2/p_{t 0}^2) \propto \ln(s/s_0)$ as a
result of the diffusion in the space of transverse
momenta\cite{BFKL}.

\par Within the double logarithmic  approximation we  evaluate analytically both the maximum in
 the  distribution over the invariant masses of the  q$\bar q$ pair which contribute to the transverse
  and logitudinal total cross section of DIS, and the corresponding average transverse momenta squared.
\par Consider first the case of the longitudinal photons. Then
   the
position of the maximum  increases with energy as \beq M^2=
M_{1L}^2(s/s_0)^{\alpha_s(N_c/\pi)/9}. \label{infu1} \eeq
Eq.\ref{infu1} is derived in the approximation $Q^2\ll M^2\ll s$
which is self-consistent at sufficiently high energies.  One can
see from this expression that  that transverse momenta of quarks
increase with the energy since $M^2=p_t^2/z(1-z)$, and the
configurations with $z=1/2$ dominate at sufficiently high
energies.
\par The dependence of the average quark transverse
momenta on energy is calculated below numerically within the
double logarithmic approximation and/or within the LO DGLAP and
BFKL approximations. For certainty we define average transverse
momentum of quark as corresponding to the median of integral for
the  total cross section. Within the double logarithmic
approximation to the cross section initiated by longitudinal
photon we obtain:
\begin{equation}
M^2_L\sim 4p_{t}^2\sim
0.7Q^2\exp(0.17((\alpha_sN_c/\pi)\log(x_0/x))^{0.55}).
\label{transverse1}
\end{equation}
Here  $x_0\sim 0.01$. The analysis was done in the interval
$s=10^4$ to $s=10^{11}$ Gev$^2$. Note that the derived rate of the
increase with the energy of the characteristic scale does not
depend on the external virtuality $Q^2$.  However, $M^2_0$ depends
on  the normalisation point in   $x_0$ and $Q_0^2$.
It is worth emphasizing
 that since we are interested here in the proof of the rise
of the transverse momenta in the current fragmentation region, we
carry for the illustration, the calculations over a very wide
spectrum of energies $s\sim 10^4 \div 10^{11}$ GeV$^2$. The detailed calculations
for the realistic energies have been carried in the
LO approximation using the CTEQ5L  gluon pdfs
\cite{BFS1,BFS2}.  Qualitatively they produce similar results
although depended on chosen extrapolation to small $x$. In
particular  the CTEQ6L parametrization leads to a significant suppression of the effects discussed in the paper.

\par Similar results were obtained for the transverse photons. In this case we were able to carry out
 an analytical calculation for the invariant mass distribution maximum for the symmetric
configurations and found that it rapidly increases with energy:
\beq M^2_{1T}\sim M_0^2 (s/s_0))^{\alpha_s(N_c/\pi)/4}.
\label{t23} \eeq The analytical results has been obtained in the
kinematics: $Q^2\ll M^2 \ll s$. It is well known however that in
the case of the transverse photons a major role in a wide
kinematical  region is played by $q\bar q$ configurations where
one of the partons carries most of the plus component of the
photon momentum.  With increase of the energy the role of
asymmetric configurations is reduced since their contribution
grows with energy more slowly. In order to take into account the
asymmetric configurations we have made a numerical calculation of
a transverse cross-section in the interval $s=10^4-10^{11}$
GeV$^2$, and obtained: \beq M^2_T\sim
0.7Q^2(x_0/x)^{0.4\alpha_sN_c/\pi}, \label{fit} \eeq $x_0\sim
0.01$.

\par If we take into account the increase of the transverse momenta
of the dipole $p_t^2$ with energy within the framework of the
dipole model and the $k_t$ factorization theorem we are lead to
the generalization of the DGLAP \cite{DGLAP} and BFKL \cite{BFKL}
approximations which is done in the paper within the LO
approximation.
\par
The rapid increase of the characteristic transverse scales in the
fragmentation region has been found first in ref. \cite{FGMS,FSW,
Rogers1,Rogers2}, within the black disk (BD)  regime. Our new
result is the prediction of the increase with energy of the jet
transverse momenta in the fragmentation region, in
 the kinematical domain where methods  of pQCD are still applicable. This effect could be considered as
  a precursor of BD regime indicating the possibility of smooth matching between two regimes.

\par
As the application of obtained results we  obtain that in pQCD
\begin{equation}
\sigma_L(x,Q^2)/\sigma_T(x,Q^2)\propto (Q^2/4p_t^2)\propto
(Q^2/s)^{\lambda}.
\end{equation}
i.e. this ratio should decrease as the  power of energy instead of being $O(\alpha_s)$.

\par
The increase of the parton momenta in the DIS in the current
fragmentation region leads to the change of many characteristics
of high energy processes. We  find that the coherence length of
the DIS processes increases with energy within pQCD as
\begin{equation}
\propto (1/2m_N)(s/Q^2)^{1-\lambda},
\end{equation}
 i.e. slower than in the parton model ($1/2m_Nx$ - the Ioffe
length).  This is the because the coherence length for a given process follows from uncertainty principle:
\beq
 l_c=(s/2m_N)/(M^2(s)+Q^2),
\label{ioffe1} \eeq where $M^2(s)\propto p_t^2(s)$ is the typical
$M^2$ important in the wave function of photon in the target rest
frame and $p_t$ is the transverse momentum of constituents in the
wave function of photon. This result has the implication for the
space structure of the wave packet describing a rapid hadron. In
the classical multiperipheral picture of Gribov a hadron has a
shape of a pancake of the longitudinal size $1/\mu$ (where $\mu $ is the soft scale)  which does
 not depend on the incident energy
\cite{Gribovspace-time}. On the contrary, we find
the biconcave shape for the rapid hadron in pQCD with the minimal
longitudinal length for small impact parameter $b$ decreasing with
increase of energy and being smaller for nuclei than for the
nucleons.

The paper  is organized in the following way. In section 2 using
the technique first introduced in QED by V.Gribov \cite{Gtrick},
we rewrite the formulae of the dipole model for the inelastic
cross-section of DIS in the form of the spectral representation
over invariant masses for both longitudinal and transverse
photons. $k_t$ factorization \cite{CH, CE} is explicitly fulfilled
in this representation.

The analysis of these formulae predicts increase with energy of transverse quark  momenta in the current
 fragmentation region.  In   section 3  we use the double logarithmic approximation for the amplitude for the
 interaction of quark dipole with the target, to
evaluate the increase with the energy of  the quark transverse
momenta in the current fragmentation region. In  section 4 we
study the dependence of coherence length on the collision energy.
In the section 5 we explain  that in pQCD rapid hadrons and nuclei
look like bi-concave lenses. Finally, in  section 6
  we discuss the possible applications of our results to pp, pA collisions at the LHC.

\section{The target rest frame description.}
\par
Within the LO approximation the QCD factorization theorem allows
to calculate the total cross section of the longitudinally
polarized strongly virtual photon scattering off a hadron target
through the convolution of the virtual photon wave function
calculated in the dipole approximation and the cross section of
the dipole scattering off a hadron. In the target rest frame the
cross section for the scattering of longitudinally polarized
photon has the form \cite {BFGMS,FS,FRS,Muellereikonal}:
\begin{equation}
\sigma(\gamma_L^{*}+T\to X)={e^2\over 12\pi^2}
\alpha_s \int d^2 p_t dz \left<\psi_{\gamma_L^{*}}(p_t,z)\right|
 \sigma(s,p_t^2) \left|\psi_{\gamma_L^{*}}(p_t,z)\right>.
\label{crosssection1}
\end{equation}
Here $\sigma$ is the dipole cross-section operator:
\begin{equation}
\sigma=(4\pi^2/3)\alpha_s(p_t^2)(-\Delta ) \cdot xG(\tilde{x}=
(M^2+Q^2)/s, M^2)), \label{cross section}
\end{equation}
 $\Delta$ is the  two dimensional Laplace operator in the space of
the transverse momenta, and
\begin{equation}
M^2=(p_t^2+m_q^2)/z(1-z), \label{M}
\end{equation}
is the invaiant mass squared of the dipole.In the coordinate
representation $\sigma$ is just a multiplication, but not a
differential operator. In the leading $\ln(x_0/x)$ approximation a
similar equation arises where the cross section is expressed in
terms of convolution of impact factor and unintegrated gluon
density.  In practice, both equations should give close results.
Using an integration over parts over $p_t$ it is easy to rewrite
Eq. ~\ref{crosssection1} within the LO accuracy in the form where
integrand will be explicitly positive:
\begin{equation}
\sigma(\gamma_L^{*}+T\to X)={e^2\over 12\pi^2} \int
\alpha_s  d^2 p_t dz  \left< \nabla{\psi_{\gamma_L^{*}}}(p_t,z)\right| f(s,z,p_t^2)
\left|\nabla \psi_{\gamma_L^{*}}(p_t,z)\right>,
\label{crosssection2}
\end{equation}
where
\begin{equation}
f=(4\pi^2/3)\alpha_s(p_t^2)
xG(\tilde{x}, M^2).
\end{equation}
In the derivation we use  boundary conditions that photon wave
function is  negligible at $p_t^2\to \infty$ and that the
contribution of small $p_t$ is the higher twist effect.

\par
The cross section of the interaction of the longitudinal photon can be rewritten in the form of spectral
 representation  by  explicitly
differentiating the photon wave function :
 \beq
\sigma_L=8\pi^2\frac{\pi\alpha_{\rm e.m.}\sum e^2_qF^2Q^2}{12}
\int dM^2 \alpha_s (M^2/4) \frac{M^2}{(M^2+Q^2)^4}\cdot
g(\tilde{x},M^2). \label{r10} \eeq Here $F^2=4/3$ for the
colorless dipoles build of color triplet constituents, and
$F^2=9/4$ for the gluonic dipoles.

The spectral representation of the electro-production amplitude
over $M^2$  is a general property of  a quantum field theory at
large energies where the coherence length significantly exceeds
the radius of  the target T  \cite{Gribov,Yennie}. The pQCD
guarantees additional general property: the smaller size  of the
configuration in the wave function of projectile photon leads to
the smaller interaction with the target but this interaction  more
rapidly increases with the energy. In the NLO approximation the
structure of formulae should be the same except the appearance of
the additional $q\bar q g, ...$ components in the wave function of
photon due to the necessity to take into account  the QCD
evolution of the photon wave function \cite{FS}.

\par
The similar derivation can be made for the scattering of the
spatially small transverse photon. In this case the contribution
of small $p_t$ region (Aligned Jet Model contribution) is
comparable to the pQCD one.  To suppress AJM contribution we
restrict ourselves in the paper by the region of large $p_t^2$ and
sufficiently small $\tilde{x}$  where pQCD contribution dominates
because of the rapid increase of the gluon distribution with the
decrease of $x$.

The pQCD contribution into the total cross-section
initiated by the transverse photon  has the form:
 \begin{eqnarray}
 \sigma_T&=& \pi^2\frac{\pi\alpha_{\rm e.m.}\sum
e^2_q F^2}{12} \nonumber\\[10pt]
&\times &\int^1_0dz \int
dM^2\alpha_s(M^2z(1-z))\frac{z^2+(1-z)^2}{z(1-z)}
\frac{(M^4+Q^4)}{(M^2+Q^2)^4}\cdot g(\tilde{x},
4M^2z(1-z)).\nonumber\\[10pt]
\label{r11}
\end{eqnarray}
Here while doing the actual calculations we introduced a cut-off
in the space of transverse of transverse momenta $M^2z(1-z)\ge u,
u\sim 0.3$ GeV$^2$.
\section{The double logaritmic approximation.}
\par
In this section we analyze the new properties of the pQCD regime
within the double log  approximation. The advantage of this
approximation is that it will enable us  to perform some of the
calculations analytically/semianalytically. We will present the
detailed numerical results for the  pattern observed in the
complete LO/NLO in a later  more detailed publication where we
will demonstrate that qualitative though not numerically the
results are the same (see however some preliminary results in LO
below).
 \par We find that at
sufficiently large energies the characteristic invariant mass of
the system of constituents
 produced by the electromagnetic current is not $Q^2$ but much larger-$M^2(s,Q^2)$. Thus in the calculations of
 of the  high energy processes the effective virtuality is $M^2$  but not $Q^2$.

\par
In the double logarithmic approximation the structure functions  are given by \cite{DDT}
\beq
g(x,Q^2)=\int dj/(2\pi i)(x/x_0)^{j-1}(Q^2/Q^2_0)^{\gamma (j)},
\label{2.10}
\eeq
where  the anomalous dimension is
$$\gamma (j)=\frac{\alpha_s N_c}{\pi (j-1)}.$$
To simplify the calculation we assume, the initial condition for the  evolution with $Q^2$:
 \beq
g(x, Q_0^2)=\delta (x-1).
\label{KS1}
\eeq
In the saddle point approximation one finds \cite{DDT}:
\beq
g(x,Q^2)=\frac{\log(Q^2/Q^2_0)^{1/4}}{\log(x_0/x)^{3/4}}\exp{\sqrt{4\alpha_s(Q_0^2)N_c/\pi\log(Q^2/Q_0^2)\log(x_0/x)}}
\label{2.11b}
\eeq
Structure function of a hadron is given by the convolution of this kernel with nonperturbative
structure function of hadron in the normalization point $Q^2=Q_0^2$.

In the following we shall neglect the pre-exponential factor, since absolute value of g as well as the
pre-exponential factor weakly influence the transverse scale, and its evolution with energy:
\beq
g(x,Q^2)=\exp{\sqrt{4\alpha_s(Q_0^2) N_c/\pi
\log(Q^2/Q_{0}^2)\log(x_0/x)}}.
\label{2.11}
\eeq

\subsection{Energy dependence of the quark
transverse momenta for fragmentation processes initiated by
longitudinal photon.}
\par
We shall find analytically the scale of the transverse momenta in
the limit  when $s\gg M^2\gg Q^2$ . For certainty we restrict
ourselves to the contribution of light quarks.
\par
At large $Q^2$ the cross-section for the scattering of the longitudinal photon is dominated by the contribution of
the spatially small dipoles, so it is legitimate to neglect the quark masses.  In this limit the
cross section is proportional to \beq \sigma_L \propto Q^2 \int
dM^2 n(M^2,s,Q^2), \label{dens} \eeq where the function
$n(M^2,s,Q^2)$
is given by  Eqs.~\ref{r10}, \ref{r11}:
\begin{eqnarray}
n(M^2,s,Q^2)&=& \alpha_s(M^2/4) \frac{M^2}{(M^2+Q^2)^4}
\nonumber\\[10pt]&\times&
\exp(\sqrt{4\alpha_s(Q_0^2)(N_c/\pi)\ln(M^2/M_0^2)\ln((s/(M^2+Q^2)((M^2_0+Q^2)/s_0))},\nonumber\\[10pt]
\label{2.11a}
\end{eqnarray}
Here we keep only large terms depending on $M^2$ (we do not write
here explicitly the $M^2$ independent overall normalization factor
irrelevant for the calculations below).

\par
Let us show that the maximum of $n(M^2, s, Q^2)$  increases with the energy. At very high energies $n$ is proportional
 to
 \begin{eqnarray}
n&\sim& \exp(\ln\alpha_s(M^2/4)+\log(M^2/Q^2)-4\log((Q^2+M^2)/Q^2)
\nonumber\\[10pt]
&+&\sqrt{4\alpha_s(N_c/\pi)(\ln(s/s_0)-\log((Q^2+M^2)/(Q^2+M_0^2))
\ln(M^2/M_0^2)}).\nonumber\\[10pt] \label{hen1} \end{eqnarray}
 In the limit of
fixed $Q^2$ but very large energies,
 $\log(s/s_0)\gg \log((Q^2+M^2)/(Q^2+M_0^2))$.  Let us assume that for the maximum:  $M^2\gg Q^2$.
  We can find the maximum of the expression \ref{hen1} under this  assumption analytically, and then
   check that this assumption is indeed self-consistent. As a result we can rewrite Eq.~\ref{hen1} as
\beq n\sim
\exp(\ln(\alpha_s(M^2/4)-3\log(M^2/Q^2)+\sqrt{4\alpha_s(Q_0^2)
(N_c/\pi)(\ln(s/s_0)\ln(M^2/M_0^2)}). \label{hen2} \eeq
Differentiating  the argument of the exponent over
$\log(M^2/M_0^2)$ we obtain the equation for the maximum: \beq
1/\ln(M^2/4M_0^2)+ 3=(1/2)(1/\ln(M^2/M_0^2) \sqrt{4\alpha_s(Q_0^2)
(N_c/\pi)\ln(s/s_0)/\ln(M^2/M_0^2)}. \label{2.12a} \eeq
\par
Neglecting the small first term we find: \beq M^2=
M_0^2(s/s_0)^{\alpha_s(N_c/\pi)/9}. \label{infu} \eeq
\par
Here $M_0^2\sim Q^2$ and $s_0\sim Q^2$. We
will refer to this extremum value of $M^2$ as  $M^2_1$.
\par
At the  extremum
  $n\propto
(\alpha_s(M_{1}^2/4)/M_{1}^6) exp
((N_c/\pi)(\alpha_s/3)\ln(s/s_0))$. Therefore \beq
\frac{d\sigma_L}{dM^2}\vert_{M^2=M^2_1}\approx
\alpha_s(M_1^2/4)(Q^2/M_1^6)(\exp((N_c/\pi)(\alpha_s(Q_0^2)/3)\ln(s/s_0))
\label{red4} \eeq
\par
However, the position of the maximum of the integrand is
not sufficient to characterize the relevant transverse scales as
a large range of $M^2$ is important in  the integrand.
(Calculation of second derivative shows that dispersion over M$^2$
is large.The width of the distribution over $\ln(M^2/M_0^2)$ is
$\sqrt (2/3)\log(M^2/M_0^2)$.) Hence we need to determine $M^2$ range which gives most of
the integrand support. For certainty, we define the  range of $M^2 \le M^2_t$ which provides a fixed,
 say, $50\%$ fraction of the total perturbative cross-section.  Let us estimate how this scale increases with
  the energy in the double log approximation. First,  let us consider the total
  cross section. The upper limit $u$ of integration over $M^2$ is
determined by the condition $M^2\ll s$.

For certainty we choose upper limit of integration as \beq M^2\le
M^2_{\ max}=0.2s, \label{m1} \eeq although the result of numerical
calculations is  insensitive to the upper bound because essential
$M^2$ are significantly smaller.
\par
Let us first calculate the median scale semianalytically. Within the double logarithmic
approximation, and assuming that the conditions
$\log(s/s_0)\gg \log((Q^2+M^2)/(Q^2+M_0^2)$, is still valid for the relevant $M^2$, the integral for the
 cross section can be written
similar to Eq.\ref{hen2}:
\begin{eqnarray} \sigma (u)&=&
(Q^2/M_0^4)\int^{\log(u/M_0^2)}_0d\ln(M^2/M^2_0)
\alpha_s(M^2/4)\exp(-2\ln(M^2/M_0^2)\nonumber\\[10pt]
&+&\sqrt{(4\alpha_sN_c/\pi)\ln(M^2/M^2_0)\ln(s/s_0)}).\nonumber\\[10pt]
\label{est}
\end{eqnarray} Here $u$ is the upper cut-off in the invariant
masses. Introducing the  new variable $t=\log(M^2/M^2_0)$, we obtain:
 \beq
 \sigma(u)=(Q^2/M_0^4)\int_0^{\kappa (u)} dt
\alpha_s(tM_0^2/4)\exp(-2t+\sqrt{(4\alpha_sN_c/\pi
)\log(s/s_0)t}), \label{s1} \eeq where $\kappa (u)=\ln(u/s_0)$.
The integral for the total cross-section is given by the equation
similar to Eq. \ref{s1}, with the upper integration limit being
 replaced by $\kappa(s)=\sqrt{\ln(0.2s/s_0)}$.
The integral \ref{s1} is actually the  error function \cite{AS},
which can be easily evaluated numerically.  Requiring that it
 gives one half of the  cross  section we find
  \beq
M^2_t\sim M_0^2(s/s_0)^{0.28\alpha_sN_c/\pi}. \label{1} \eeq

Evidently, for sufficiently large $s$ our initial assumption
$\log(M^2/M_0^2)\gg \log(Q^2/Q_0^2)$ is fully self-consistent.
This is because the decrease of $n$ with $M^2$ due to $1/M^6$
terms in the integrand of  Eq. \ref{r10} is partially compensated
by the rising exponential, giving a relatively slow decrease of
$n$ to the right of its maximum.
\par
Note that the rate of the increase of $M^2_t$ with $s$ is much
higher than for $M^2_1$ due to the slow decrease of the integrand
with $M^2$. The cross section of jet production with $M^2$ at this interval also
increases with the energy as
 \beq \frac{d\sigma}{dM^2}\vert_{M^2=M^2_{\rm T}}\sim (s/s_0)^{0.24\alpha_sN_c/\pi}.\label{red301}\eeq
\par  In order to understand the dependence of the median scale on both the energy and $Q^2$ quantatively
we made numerical calculation of the characteristic transverse momenta  using the DGLAP double log structure function.
We find that the increase rate of the transverse
 momenta indeed does not depend on the external
virtuality $Q^2$. Considering the wide interval of energies and
$s=10^4\div 10^{11}$ GeV$^2$, and $20<Q^2<200$ GeV$^2$ we obtain
the approximate formulae: \beq M^2_t\sim
0.7Q^2)\exp(0.17((4\alpha_sN_c/\pi
)\log(x_0/x))^{0.55}).\label{transverse11}\eeq We give this
estimate only for illustrative purposes, since the double
logarithmic approximation is  semirealistic only. Still our
results indicate that for external virtualities $Q^2<100 $ GeV$^2$
and energies which can be reached at LHeC the onset of a new pQCD
regime may take place.

The cross section of the jet production at this scale also
increases with the energy as
 \beq \frac{d\sigma}{dM^2}\vert_{M^2=M^2_{\rm t}}\sim \frac{1}{Q^4}\frac{f(x)}{1+0.7f(x)}
 G(x(1+0.7f(x),0.7Q^2f(x),\label{red3}\eeq
where \beq f(x)=\exp(0.17((4\alpha_sN_c/\pi
)\log(x_0/x))^{0.55}).\label{f}\eeq

\subsection{Transverse photon: the characteristic transverse scale in the photon fragmentation region.}

\par It is well known that the main difference between the longitudinal and transverse structure functions
in the DIS is the presence of the strongly asymmetrical in $z$
configurations due to the presence of the $(z(1-z))^{-1}$
multiplier in the spectral density. As a result there is a
competition between two effects.  One is  a slower decrease of the
spectral function with $M^2$ (by the factor $M^2/Q^2$), leading to
the  more rapid increase of the characteristic transverse momenta
for the symmetric configurations. The second effect is presence of
the asymmetric ($z\to 0$) configurations which are  characterized
by the small transverse momenta $k^2_t$ for a given invariant mass
$M^2$. For such configurations the rate of increase of the gluon
structure function with energy is small.
\par Let us first show that the transverse momenta increase rapidly
for symmetric configurations.
The spectral representation for the transverse photon for
symmetric configurations is proportional to \begin{eqnarray}
n(M^2,Q^2,s)&\sim& \frac{M^4+Q^4}{(M^2+Q^2)^4}\nonumber\\[10pt]
&\times&\sqrt{4\alpha_s(N_c/\pi)(\ln(s/s_0)-\log((Q^2+M^2)/(Q^2+M_0^2))
\ln(M^2/M_0^2)}).\nonumber\\[10pt] \label{t1} \end{eqnarray}
Similar to the case of the longitudinal photon we obtain   for high energies, when
$M^2_1\gg Q^2$, the dependence of the maximum of $n$ on  energy: \beq
M^2_1\sim M_0^2(s/s_0))^{\alpha_s(N_c/\pi)/4}, \label{t2} \eeq
i.e. the increase rate is twice as fast as compared to the case of
longitudinal photon. $M^2_1$ increases with $s$ at high energies and
thus the condition $M^2\gg Q^2$ is perfectly self-consistent at
very high energies.
\par The jet  cross section at the maximum of the curve also
increases as \beq \frac{d\sigma}{dM^2}\vert_{M^2=M^2_1}\sim
(s/s_0)^{\alpha_sN_c/2\pi}\label{red2}\eeq
\par In addition we calculate the total cross-section in the same
approximation semi-analytically getting the error function and
obtain the rate of increase $(s/s_0))^{0.14(4\alpha_sN_c/\pi)}$,
which is twice that for
 the longitudinal case.
\par However, as we mentioned above, a considerable contribution of the nonsymmetrical configurations has
 the opposite effect. In order to take these configurations into account we performed a numerical calculation
 using the gluon
distribution function within the double log approximation. The
result is that the characteristic median scale $M^2_t$ increase
like \beq
M^2_0(s/s_0)^{0.1(4\alpha_sN_c/\pi)}\label{transverse21}\eeq The
value of the exponent is $0.12$ for the beginning of the studied
energy range  $s\sim 10^4 \div 10^{11}$ GeV$^2$, and decreases to
0.09 at the upper end (for typical $\alpha_s=0.25$. Thus the rate
of the increase with the energy is approximately the same as for
longitudinal photons for not very high energies. For very high
energies the symmetric configurations win over asymmetric ones,
leading to twice as rapid increase of the transverse momenta than
in the longitudinal case.
\par
The precise determination of the scale $M_0^2(Q^2)$ is beyond the
accuracy of this paper. Effectively we obtain the dependence
$M^2_t\sim 0.7Q^2 (x_0/x)^{0.1(4\alpha_sN_c/\pi)}$.
\par
One can also estimate the rate of the increase of the jet
production cross section : \beq d\sigma_T/dM^2_{\rm M^2_t}\sim
\frac{1}{Q^4}\frac{1+0.5h(x)^2}{(1+0.7h(x))^4}G(x(1+h(x),Q^2(1+h(x)),
\label{density3}\eeq \beq
h(x)=(x_0/x)^{0.1(4\alpha_sN_c/\pi)}\label{df23}\eeq We found a
rapid increase of the jets multiplicity. Thus the rate of the
increase with energy of the transverse momenta of quarks in the
current fragmentation region for transversely polarized photon is
significantly more rapid. Consequently we find that
$\sigma_L/\sigma_T \approx Q^2/M^2$ being numerically small should
slowly decrease with energy at sufficiently high energies .
\par We conclude that it is possible to show analytically that for
very high (asymptotic) energies the relevant invariant masses
extend well  beyond $Q^2$ and increase with the energy.
\par
The direct numerical calculation of the $M^2_t$ scale shows that the rate of increase
is independent of external virtuality.
\subsection{The leading logarithmic approximation.}

\par
The above results were obtained in the double logarithmic
approximation. It is also possible to carry out the numerical
calculation in LO approximation using the CTEQ5L gluon
distribution functions \cite{BFS}. In this approximation the
median scale still increases as $M^2_t\sim 0.7Q^2
(x_0/x)^\lambda$, where $x_0\sim 10^{-2}$, and $\lambda\sim 0.06$
for longitudinal and $\lambda\sim 0.08$ for transverse photons.
Although this increase is quite slow, the rise of momenta is not
negligible: for energy increase from $10^4$ to $10^7$ GeV$^2$ the
scale increases by a factor $\sim 1.5$. The use of CTEQ6L will
decrease the considered effects.
\section{The coherence  length.}
\par
 In the previous sections we determined the  energy dependency of the effective transverse scale
  at high energies which allowed us to
evaluate coherence length.The coherence length $l_c$ corresponds
to the life-time of the dipole fluctuation at a given
energy in the rest frame of the target. The original suggestion of the existence of the coherence
length in the deep  inelastic scattering was first made by Ioffe, Gribov and Pomeranchuk \cite{GIP,Gribov}.
It was found already in the sixties within the parton model approximation by  Ioffe \cite{Ioffe} that
 the coherence length at moderate $x_B$   is $l_c\sim 1/2m_Nx_B$
i.e. it linearly increases with energies. In pQCD coherence length
\begin{equation}
L_c=(1/2m_Nx)(s_0/s)^{\lambda}
\end{equation}
Less rapid increase of $L_c$ with energy has been found before
in the numerical calculations of structure functions in the target rest frame accounting for $Q^2$ evolution of structure functions .
\cite{KS,BFcoh}

It is worth noting that the discussed pattern of  the energy dependence of the coherence length leads to a change
 of the structure of the fast hadron wave function as compared to the Gribov picture
\cite{Gribovspace-time} where the longitudinal size of the hadron is determined by the wee parton cloud and energy
 independent
$L_z\sim 1/\mu$. Here $\mu\sim \, .3 \div 0.4 \, GeV/c$  is the
soft mass scale. On the other hand a slower than $1/m_Nx$ rate of
the increase of the coherent length with energy leads to a
decrease of the longitudinal size of the hadron with energy. The
typical size is determined by  the BD momentum at a given impact
parameter for a particular energy. Moreover  since the BD momentum
is larger for small impact parameters the
 nucleon has a form of
a double concave lens. It is of interest also that for the zero
impact parameter the longitudinal size of a  heavy nucleus is
smaller than for a nucleon.

\section{The form of nucleon, nucleus in DIS}
\par Our results have  the important consequences for the
transverse  structure of the hadrons and nuclei.
\par
Let us consider the longitudinal distribution of the partons in a
fast hadron. As it was already mentioned in the previous section,
in the parton model the longitudinal spread of gluonic cloud is
$L_z\sim 1/\mu$ for the wee partons and is much smaller than for
harder partons, with $L_z \sim 1/xP_h$ for partons carrying a
finite $x$
 fraction of the hadron momentum \cite{Gribovspace-time}. The picture is changed qualitatively in the
 limit of very high energies when interactions
reach BD regime for $k_t \gg \mu$. In this case the smallest
possible characteristic momenta in the frame where hadron is fast
are of the order $k_t (BDR)$ which is a function of both initial
energy and transverse coordinate, $b$ of the hadron.
Correspondingly, the longitudinal size is $\sim 1/ k_t (BDR) \ll
1/\mu$. Since the gluon parton density  decreases with the
increase of $b$ the longitudinal size of the hadron is larger for
large $b$, so a hadron has a shape of biconcave lens, see Figs. 1,
2.
\par
We depict the typical transverse structure of the fast
nucleon in Fig.~1.  We see that it is drastically
different from the naive picture of a fast moving nucleon as a flat narrow disk with small
constant thickness.
\par
Consider now the case of the DIS on the nuclei.
 \par
Consider first the case of external virtualities of the order of several GeV. In this case the
 shadowing effects mostly  cancel the $A^{1/3}$ for a given  impact parameter, $b$ \cite{FS} and
 the gluon density in the nuclei is comparable to that in a single nucleon for $b\sim 0$.
Consequently over the large range of the impact parameters the
nucleus longitudinal size is approximately the
 same as in the nucleon at $b\sim 0$.
\par Consider now the case of the large external virtualities
$Q^2 \ge  40$ GeV$^2$. In this case the leading twist
 shadowing is small,  and the corresponding gluon density
unintegrated over b is given by a product of a nucleon gluon
density and the nuclear profile function: \beq T(b)=\int dz
\rho(b,z),\label{bor1}\eeq where the nuclear three-dimensional
density is normalized to A. We use standard Fermi step
parametrization  \cite{Bohr}
 \beq \rho
(r)=C(A)\frac{A}{1+\exp((r-R)/a}, R=1.1A^{1/3} {\rm fermi}, a=0.56
{\rm fermi}.\label{improved}\eeq Here $r=\sqrt{z^2+b^2}$,
 and A is the atomic number.
C(A) is a  normalization factor, that can be
calculated numerically from the condition
$\int d^3r \rho(r)=A$.  At the zero
impact parameter $T(b) \approx 0.5 A^{1/3}$ for large A.
\par The dependence of the thickness of
a fast  nucleus as a function of
the transverse size is depicted in Fig.~2 for a typical high
energy $s=10^7$ GeV$^2$, $Q^2=40$ GeV$^2$. We see that the nuclei also has a form of a biconcave
lens instead of a flat disk.
\par Note that this picture is very counterintuitive. We see that the nuclei is
thinner than a single nucleon, i.e. the thikness of a nuclei is
smaller than for a nucleon although we have $\sim A^{1/3}$
nucleons at the central impact parameter. In other words the
longitudinal extend  of a
system is smaller than the the longitudinal extend of its
constituents. The explanation of this BDR phenomenon is
straightforward. For a given impact
parameter $b $, the longitudinal size of a heavy nucleus $1/ k_t
^{(A)}(BDR) < 1/ k_t ^{(p)}(BDR)$ since the gluon distribution
function in the nuclei $G_A(x,b) > G_N(x,b)$. So a naive classical
picture of a system build of the constituents being larger than
each of the constituents is grossly violated. This situation is in
some respects analogous to the phenomenon of color
transparency/existence of the point-like configurations). Thus we
have a paradox: in the fast reference frame the nuclei is much
thinner than any of its constituents.

\begin{figure}[ht]
\centering\includegraphics[height=10cm,width=14cm]{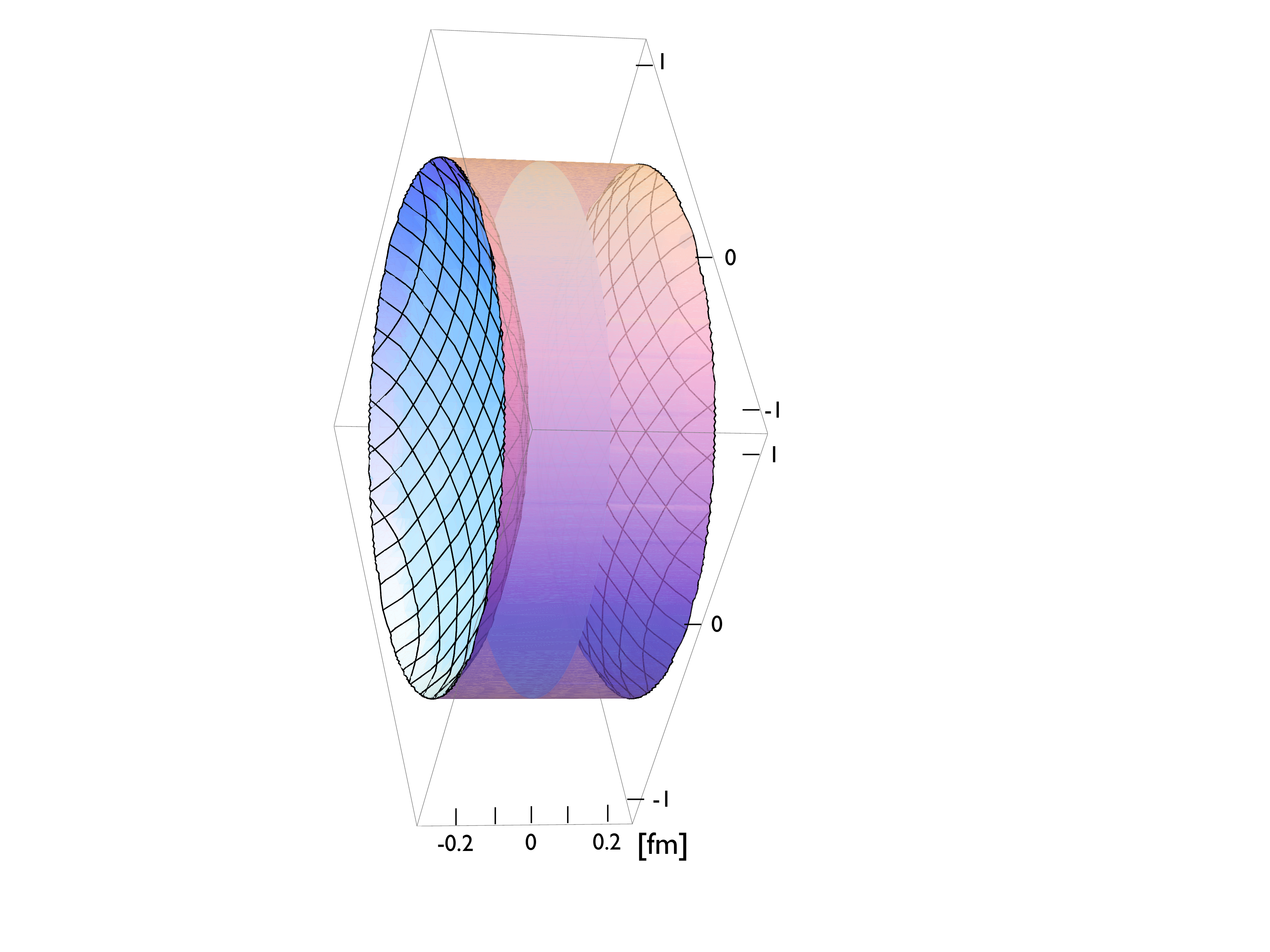}
\caption{3D image of the fast nucleon at
$s=10^7  \, \mbox{GeV}^2 $ and $Q^2=40 \mbox{GeV}^2$.}
\label{f1}
\end{figure}

\begin{figure}[ht]
\centering\includegraphics[height=10cm,width=14cm]{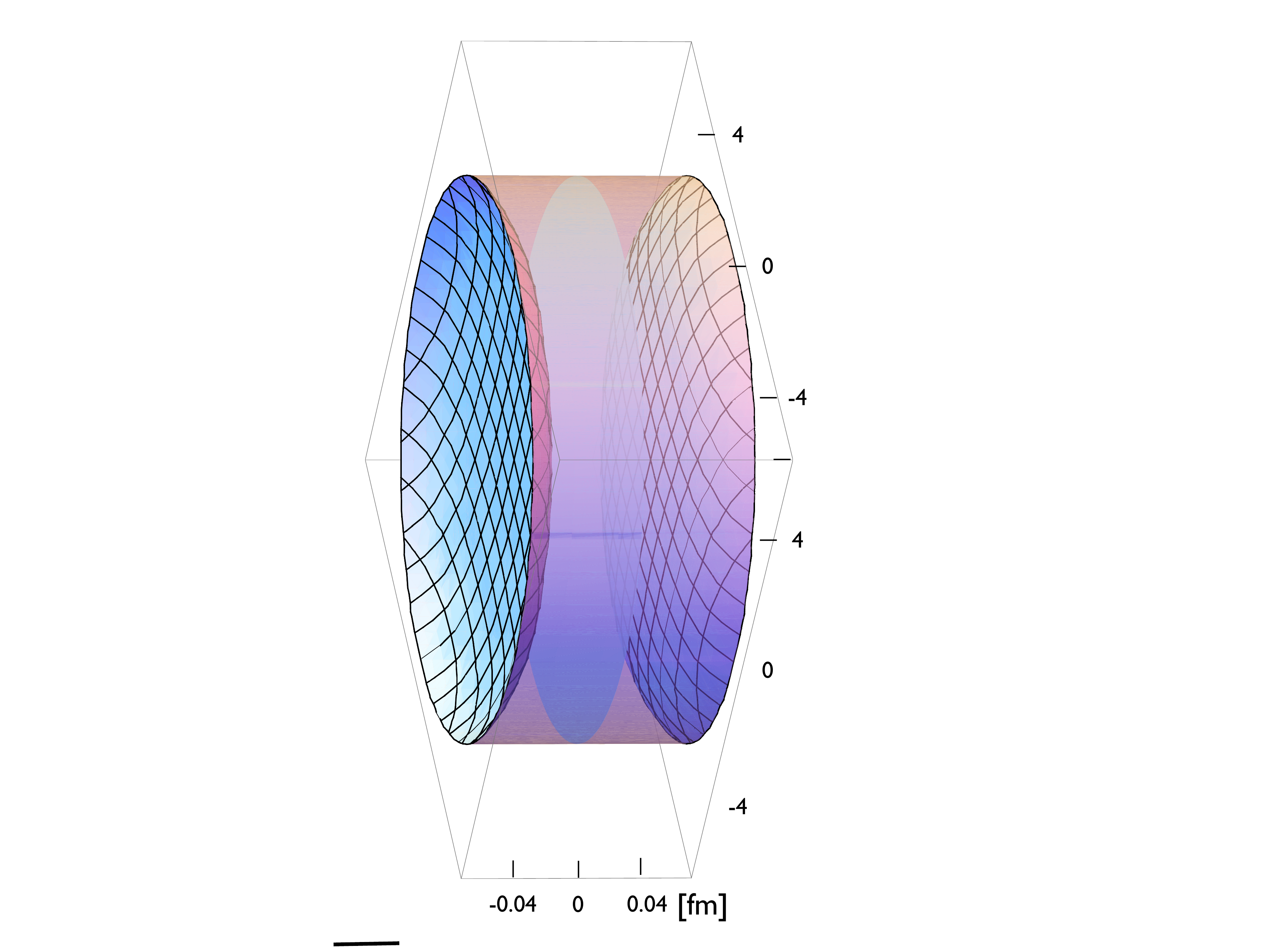}
\caption{3D image of the fast heavy nucleus (gold) at
$s=10^7  \, \mbox{GeV}^2 $ and $Q^2=40 \mbox{GeV}^2$.}
\label{f2}
\end{figure}

\par
The resolution of the paradox in the BD regime is quite simple:
the soft fields of individual nucleons destructively interfere
cancelling each other.  So a naive classical picture of a system
build of the constituents being larger than each of the
constituents is grossly violated.

\section{Experimental consequences.}

\par
The current calculations of  cross-sections of hard processes at the LHC are based on the
use of the DGLAP parton distributions and the
application of the factorization theorem. Our results imply that
the further analysis is needed to define the kinematic regions
where one can use DGLAP distributions. We showed in the paper
that for DIS at high energies there are kinematic regions where one is forced to use a $k_t$
factorization and the dipole model instead of
the direct use of DGLAP. A similar analysis must be made for the
$pp$ collisions at LHC.
The expected effect is the increase with energy of the probability
of the small dipoles in the wave function of proton
\cite{Brodsky}. Quantitative analysis of this problem will be
presented elsewhere.
\par
The hard processes initiated by the real photon can be directly
observed in the ultrapheripheral collisions \cite{Mark1}. The
 processes where a real photon scatters on a target,
 and creates two jets with an invariant mass $M^2$, can be analyzed in the dipole model by formally putting
 $Q^2=0$, while $M^2$ is an invariant mass of the jets. In this case  with a  good accuracy the spectral density
discussed above will give the spectrum of jets in the
fragmentation region. Our results show that
 the jet distribution over the transverse momenta will be broad
 with the maximum moving towards larger
transverse momenta with increase of the energy and centrality of
the $\gamma A$ collision.

\par Finally, our results can be checked directly, if and when the
LHeC facility will be built at CERN. One of us B.Blok thanks S.Brodsky for the useful discussions
of the results obtained in the paper.
This work was supported in part by the US DOE Contract Number
DE- FG02-93ER40771 and BSF.


\newpage\begin{thebibliography}{}
\bibitem{BFGMS} S. J. Brodsky, L. Frankfurt, J. F. Gunion, A. H. Mueller and M. Strikman, Phys. Rev. D50 (1994) 3134.
\bibitem{AFS}H. Abramowicz, L. Frankfurt and M. Strikman, Surveys
in High Energy Physics, 11 (1997) 51.
\bibitem{FMS} L. Frankfurt, G.A. Miller, M. Strikman, Ann. Rev.
Nucl. Part. Sci., 44 (1994) 501.
\bibitem{BBFS}B. Blaettel, G. Baym,
L. Frankfurt and M. Strikman, Phys. Rev. Lett., 70 (1993) 896.
\bibitem{Muellereikonal}  A.~H.~Mueller Nucl.\ Phys.\  B {\bf 415}, 373 (1994).
\bibitem{CFS} J.C. Collins,  L. Frankfurt and M. Strikman, Phys. Rev. D56 (1997) 2982.
\bibitem{BFKL} E. Kuraev, V. Fadin, L. Lipatov, Sov. Phys.-JEP, 44
(1976) 443; 45 (1977) 199. I. Balitsky and L. Lipatov, Sov. J.
Nucl. Phys., 28 (1978) 822.
\bibitem{DGLAP} G. Altarelli and G.Parisi, Nucl. Phys., B126
(1977) 298; V.N. Gribov and L. N. Lipatov, Sov. J. of Nucl. Phys.,
15 (1972) 438,672; Yu.L. Dokshitser, Sov. Phys. JETP 46 (1977)
641.
\bibitem{BFS1}B. Blok, L. Frankfurt and M. Strikman, in
preparation.
\bibitem{BFS2} B. Blok, L. Frankfurt and M. Strikman,
arXiv:0808.2006(hep-ph) Talk at small x conference, Columpari,
Crete,  July 2008.
\bibitem{FGMS}
 V. Guzey, L. Frankfurt, M. Strikman, M.McDermott, Eur.J. of Physics, C16 (2000) 641.
\bibitem{FSW} L. Frankfurt, M, Strikman and C. Weiss,
Ann.Rev.Nucl.Part.Sci.55 (2005) 403-465.
\bibitem{Rogers1}
T.~C.~Rogers, A.~M.~Stasto and M.~I.~Strikman,
Unitarity Constraints on Semi-hard Jet Production in Impact Parameter  Space,''arXiv:0801.0303 [hep-ph].
\bibitem{Rogers2}
  T.~C.~Rogers and M.~I.~Strikman,
  Hadronic interactions of ultra-high energy photons with protons and  light  nuclei in the dipole picture,''
  J.\ Phys.\ G {\bf 32}, 2041 (2006)
\bibitem{Gribovspace-time} V. N. Gribov, Space-time description of hadron interactions at high-energies.
 In *Moscow 1 ITEP school, v.1 'Elementary particles'*,
65,1973. e-Print: hep-ph/0006158.
\bibitem{Gtrick} V.N.Gribov
 The theory of complex angular momenta: Gribov lectures on theoretical physics. Cambridge, UK: Univ. Pr. (2003)
\bibitem{CH} S. Catani, M. Ciafaloni, F. Hauptmann, Nucl. Phys.
B366 (1991) 135.
\bibitem{CE} J. Collins, K. Ellis, Nucl. Phys., B360 (1991) 3.
\bibitem{FS} L. Frankfurt and M. Strikman, Phys. Rept., 160
(1988) 235.
\bibitem{FRS} L. Frankfurt, A.Radyushkin, M. Strikman, Phys. Rev.
D55(1997) 98.
\bibitem{Gribov} V. N. Gribov, Sov. Phys. JETP 30 (1970) 709.
\bibitem{Yennie}
T.H. Bauer, R.D. Spital, D.R. Yennie  , F.M. Pipkin,
Rev.Mod.Phys.50 (1978) 261, Erratum-ibid.51(1979) 407.
\bibitem{DDT} Yu. Dokshitzer,D. Diakonov and S. Troyan, Phys. Reports, 58  (1980) 269.
\bibitem{AS}M. Abramowitz and I. Stegun, Handbook of special
functions,Dover Publications, New York,1964.
\bibitem{BFS} B. Blok, L. Frankfurt and M. Strikman, in
preparation.
\bibitem{GIP} L. B. Ioffe, V. Gribov, I. Pomeranchuk,
Sov. J. of Nucl. Phys., 2 (1966) 549.
\bibitem{Ioffe} B. L. Ioffe, Phys. Lett., B30 (1969) 123.
\bibitem{Bohr} A. Bohr and B.R. Mottelson, Nuclear structure, v.1,
W.A. Benjamin, New York, 1969.
\bibitem{KS} Y. Kovchegov and M. Strikman, Phys. Lett., B516(2001) 314.
\bibitem{BFcoh} B.Blok and L.Frankfurt Phys.Lett.B630 (2005)
49-57.
\bibitem{Mark1} L. Frankfurt and M. Strikman, in Phys. Reports, 455 (2008) 105.
\bibitem{Brodsky} We are
indebted to S.Brodsky for emphasizing this point.
\end{thebibliography}
\end{document}